\documentclass{article}
\usepackage[utf8]{inputenc}

\title{Equivalence Testing: The Power of Bounded Adaptivity}
\author{Diptarka Chakraborty\footnote{National University of Singapore, Singapore. Email: diptarka@comp.nus.edu.sg}\and Sourav Chakraborty \footnote{Indian Statistical Institute, Kolkata. Email: sourav@isical.ac.in} \and Gunjan Kumar \footnote{National University of Singapore, Singapore. Email: dcsgunj@nus.edu.sg} \and Kuldeep S. Meel \footnote{University of Toronto, Toronto.  Email: meel@cs.toronto.edu} }
\date{}

\usepackage{amsthm,amsmath,amssymb}
\usepackage{thmtools}
\usepackage{dsfont}
\usepackage{bm}
\let\mathbbm\mathds

\newcommand{\E}{\mathbb{E}}

\usepackage{algorithm}
\usepackage[noend]{algpseudocode}
\usepackage{fullpage}
\usepackage{todonotes}
\usepackage{tikz}
\newcommand{\EstProb}{\ensuremath{\mathsf{EstProb}}}
\newcommand{\EstTail}{\ensuremath{\mathsf{EstTail}}}
\newcommand{\Accept}{\ensuremath{\mathsf{Accept}}}
\newcommand{\Reject}{\ensuremath{\mathsf{Reject}}}
\newcommand{\EquivTester}{\ensuremath{\mathsf{EquivTester}}}
\newcommand{\TP}{\ensuremath{\mathsf{TP}}}
\newcommand{\mcD}{\mathcal{D}}
\newcommand{\mcP}{\mathcal{P}}
\newcommand{\mcQ}{\mathcal{Q}}
\newcommand{\R}{\ensuremath{\mathsf{R}}}
\newcommand{\SAMP}{\ensuremath{\mathsf{SAMP}}}
\newcommand{\COND}{\ensuremath{\mathsf{COND}}}
\newcommand{\epP}{\ensuremath{\mathsf{ep}_1}}
\newcommand{\epQ}{\ensuremath{\mathsf{ep}_2}}
\newcommand{\etP}{\ensuremath{\mathsf{et}_1}}
\newcommand{\etQ}{\ensuremath{\mathsf{et}_2}}
\newcommand{\bad}{\ensuremath{\mathsf{Bad}}}
\newcommand{\Good}{\ensuremath{\mathsf{Good}}}
\newcommand{\bin}{\ensuremath{\mathsf{Bin}}}

\renewcommand{\epsilon}{\varepsilon}

\newtheorem{theorem}{Theorem}[section]
\newtheorem{claim}[theorem]{Claim}
\newtheorem{lemma}[theorem]{Lemma}
\newtheorem{definition}[theorem]{Definition}

\begin{document}

\maketitle
\thispagestyle{empty}
\setcounter{page}{0}
\begin{abstract}
  Equivalence testing, a fundamental problem 
in the field of distribution testing, 
seeks to infer if two unknown distributions
on $[n]$ are the same or far apart in the
total variation distance. Conditional
sampling has emerged as a powerful query
model and has been investigated by
theoreticians and practitioners alike,
leading to the design of optimal algorithms
albeit in a sequential setting (also
referred to as adaptive tester). 
Given the profound impact of parallel
computing over the past decades, there has been a
strong desire to design algorithms that
enable high parallelization. Despite
significant algorithmic advancements over
the last decade, parallelizable techniques
(also termed non-adaptive testers) have
$\Tilde{O}(\log^{12}n)$ query complexity, a
prohibitively large complexity to be of
practical usage. 
Therefore, the primary challenge is whether
it is possible to design algorithms that
enable high parallelization while achieving
efficient query complexity. 

Our work provides an affirmative answer to
the aforementioned challenge: we present a
highly parallelizable tester with a query
complexity of $\Tilde{O}(\log n)$, achieved
through a single round of adaptivity,
marking a significant stride towards
harmonizing parallelizability and
efficiency in equivalence testing.
\end{abstract}
\newpage

\section{Introduction}
\label{sec:intro}
Evaluating different properties of an unknown object is a fundamental challenge in statistics. When dealing with large objects, it becomes essential to determine these properties by making only a limited number of queries to the object. In the case of unknown objects being probability distributions, the goal is to assess whether the input distribution(s) possess specific properties or deviate significantly (i.e., ``$\epsilon$-far" for some $\epsilon >0$) from meeting them. All of this needs to be accomplished while minimizing the number of queries made (also known as \emph{query complexity}) to the distribution(s). Probability distributions are crucial subjects of study, and distribution testing has remained central to sublinear algorithms and modern data analysis since its introduction~\cite{goldreich1998property, goldreich2011testing, batu2013testing}.

Early investigations into distribution testing primarily utilized the $\SAMP$ query model, which only allows drawing samples from the given distribution(s). However, for testing many interesting properties, the $\SAMP$ model proves to be restrictive, as evidenced by strong polynomial (in domain size) lower bounds on the sample complexity. To overcome this limitation, several alternative query models have been proposed over the past decade. Among these models, the conditional sampling model ($ \COND $)~\cite{chakraborty2013power,crs2014testing} has been extensively studied. This model permits drawing samples from the input distribution(s) conditioned on any arbitrary subset of the domain. Various distribution testing problems have been explored under the $ \COND $ model~\cite{falahatgar2015faster, kamath2019anaconda, narayanan2021tolerant, CKM23} and certain variants of it like \emph{subcube conditioning model}~\cite{bhattacharyya2018property, canonne2021random, chen2021learning}. Moreover, the $\COND$ model and its variants have recently found applications in the areas like formal methods and machine learning (e.g.,~\cite{chakraborty2019testing, meel2020testing, golia2022scalable}).

In this work, we study the \emph{equivalence testing} problem, one of the most fundamental problems in distribution testing. In this problem, given query access to two (unknown) distributions $\mcP$ and $\mcQ$, the objective is to decide whether they are equal or $\epsilon$-far from each other in the \emph{total variation distance}. For this problem, a query-optimal algorithm is already known in the {\COND} model. However, the primary challenge with the query-efficient algorithm/tester in the $\COND$ model is its inherent sequential or adaptive nature. A tester is considered non-adaptive if it can generate all its queries based solely on the input parameter (in this case, the domain size) and its internal randomness, without relying on previous query responses. Non-adaptive testers are generally favored in practical situations because they can make multiple queries simultaneously.

Exploring the balance between the degree of adaptivity and query complexity is a captivating area of research. This curiosity prompted~\cite{canonne2018adaptivity} to delve deeper into adaptive testing, introducing a nuanced approach by permitting a limited number of adaptive stages or rounds. This multi-stage or bounded-round adaptivity concept finds resonance in various other problems, including group testing~\cite{du2000combinatorial, damaschke2013two, eberhardt2020multi}, submodular function maximization~\cite{balkanski2018adaptive, chekuri2019parallelizing}, compressed sensing and sparse recovery~\cite{nakos2018improved, kamath2019adaptive}, multi-armed bandits problem~\cite{agarwal2017learning}. In this work, we initiate the study of bounded-round adaptivity in the context of the equivalence testing problem and provide a query-efficient one-round adaptive tester.

\iffalse

We now briefly explain the previous non-adaptive algorithm~\cite{} for equivalence testing. The algorithm first selects a random power of $2$ between $2$ and $n$. Then it chooses a random set $S \subseteq [n]$ of this size. Now it performs $O(\log^6 n)$ $COND_p(S)$ and $COND_q(S)$ queries and form an empirical distribution $\hat{p}_S$ and $\hat{q}_S$ on $S$. The above procedure is repeated $O(\log^6 n)$ times. If every time the empirical distributions $\hat{p}_S$ and $\hat{q}_S$ is close in $l_{\infty}$ distance then the algorithm concludes that distributions $p$ and $q$ are the same. Hence the query complexity of their algorithm is $O(\log^{12} n)$

\fi

\section{Notations and Preliminaries}
\label{sec:prelims}
 Throughout this paper, we consider distributions over the domain $[n]:=\{1,2,\cdots,n\}$. For any $i,j \in [n]$ and $S \subseteq [n]$, for brevity we use $i \cup j$ and $i \cup S$ to denote the sets $\{i\} \cup \{j\}$ and $\{i\} \cup S$ respectively.

Given a distribution $\mcD$ over $[n]$ and $i \in [n]$, we use $\mcD(i)$ to represent the probability mass function of $i$. Similarly, for  $S \subseteq [n]$, we use $\mcD(S)$ to denote $\sum_{i \in S}\mcD(i)$. For any $\gamma \in (0,1)$, a $\pm \gamma$ estimate of a quantity (say $d$) means a number $\tilde{d} \in \left[d-\gamma , d+\gamma\right]$.

The total variation distance between two distributions $\mcP$ and $\mcQ$, denoted by $d_{TV}(\mcP,\mcQ)$ is defined as
\begin{align*}
    d_{TV}(\mcP,\mcQ) := \frac{1}{2} \sum_{i \in [n]} |\mcP(i)-\mcQ(i)|.
\end{align*}
%$$.
    
If the variation distance between two distributions is more than $\epsilon$, then we say the distributions are $\epsilon$-far (or just far, when it is clear from the context). 

The Binomial distribution, with parameters $n \in \mathbb{Z}^+$ and $p \in [0,1]$ denoted by  $\bin(n,p)$ is the distribution of the number of successes in $n$ independent experiments, where each experiment yields a Boolean outcome, with success occurring with probability $p$ and failure with probability $1-p$.

\begin{definition}[$\COND$ Query Model]
\label{def:cond}
   A conditional sampling oracle for a distribution $\mcD$ is defined as follows: the oracle takes as input a subset $S \subseteq [n]$ and returns an element $j \in S,$ such that the probability that $j \in S$ is returned is equal to $\mcD(j)/\mcD(S)$ if $\mcD(S) > 0$ and $1/|S|$ if $\mcD(S) = 0$.

   We denote such a conditional query by $\COND_{\mcD}(S).$ 
\end{definition}

The formal definition of a $k$-round adaptive tester is given in~\cite{canonne2018adaptivity}. For completeness, we present the formal definition of a one-round adaptive tester for equivalence in the $\COND$ Query Model.  

\begin{definition} 
   Given conditional query access to distributions $\mcP$ and $\mcQ$ (over domain $[n]$), and given tolerance parameter $\epsilon$ as inputs, a one-round adaptive tester $\mathcal{A}$ makes conditional queries to the distributions in two rounds:
   \begin{enumerate}
       \item In the first round, the algorithm $\mathcal{A}$ (without making any queries to the distributions)  selects a set of subsets (say $S^0_1, \dots S^0_{q_1}$ of $[n]$ and then makes 
       the conditional queries  $\COND_{\mcP}(S^0_1), \dots, \COND_{\mcP}(S^0_{q_1})$ and  $\COND_{\mcQ}(S^0_1), \dots, \COND_{\mcQ}(S^0_{q_1})$.
       \item In the second round, based on the answers to the queries it has received in the first round, it selects another set of subsets (say $S^1_1, \dots S^1_{q_2}$ of $[n]$ and then makes 
       the conditional queries  $\COND_{\mcP}(S^1_1), \dots, \COND_{\mcP}(S^1_{q_2})$ and  $\COND_{\mcQ}(S^1_1), \dots, \COND_{\mcQ}(S^1_{q_2})$.
   \end{enumerate}
   Finally, based on the answers to all the $2(q_1 + q_2)$ queries, $\mathcal{A}$ outputs with the following guarantee:
   \begin{itemize}
       \item if $\mcP$ and $\mcQ$ are identical, then with probability at least $2/3$, $\mathcal{A}$ outputs $\Accept$, and
       \item if $d_{TV}(\mcP,\mcQ) \ge \epsilon $, then with probability at least $2/3$, $\mathcal{A}$ outputs $\Reject$.
    \end{itemize}
   The query/sample complexity of the algorithm is $2(q_1 + q_2)$.
\end{definition}

In our proof, we will extensively use concentration lemmas. In particular, we will use the following version of Chernoff bound. 

\begin{lemma}[Additive Chernoff bound]\label{lem:additiveChernoffBound}
    Let $X_1,\dots,X_m$ be $m$ iid random variables, each $X_i$  takes value in $\{0,1\}$ and $\E[X_i] = p$. Then for any $\gamma \in (0,1),$
    \begin{align*}
       \Pr\left[\left|\sum_{i\in[m]} X_i/m - p \right| \ge \gamma\right] <e^{-\gamma^2m}.
    \end{align*}
\end{lemma}

%\todo{Additive Chernoff bound, round-complexity, define $\pm \delta$ estimate, define bin(n,p)}

\section{Related Work}
\label{sec:related-work}
In the standard $\SAMP$ model, the query complexity of the equivalence testing problem is $\Theta(\max(n^{2/3}/\epsilon^{4/3},\sqrt{n}/\epsilon^2))$\\~\cite{chan2014optimal,batu2013testing,valiant2011testing}, which is prohibited in most practical applications. The $\COND$ model turns out to be beneficial in this context, enabling to require only $\Tilde{O}(\log \log n)$ samples~\cite{falahatgar2015faster}, which has recently been shown to be optimal~\cite{chakraborty2024tight}.  

Unfortunately, the above-mentioned optimal tester in the {\COND} model is sequential. In simpler terms, the tester is adaptive, meaning that each query (indexed as $t$ for any $t \ge 1$) in the $\COND$ model depends on the answers to the preceding $t-1$ queries. Designing a parallel, ideally entirely non-adaptive, tester remains an enormous challenge. ~\cite{kamath2019anaconda} introduced a non-adaptive tester for the equivalence testing problem, which required $\Tilde{O}(\log^{12} n / \epsilon^2)$ queries\footnote{Note that $\Tilde{O}(f(n))$ notation hides $poly (\log f(n))$ terms.}. However, the substantial poly-logarithmic dependency on the domain size is impractical in many real-world applications. Moreover, the best-known lower bound for the query complexity of non-adaptive testers is $\Omega(\log n)$~\cite{acharya2018chasm}, indicating considerable room for improvement in the upper bound. One exciting question is to make the tester as less adaptive as possible while attaining the optimal non-adaptive query complexity. Such a question motivates the researchers to study the trade-off between the number of adaptive rounds and the query complexity (for testing various properties). The work~\cite{canonne2018adaptivity} led to the establishment of a ``hierarchy theorem" examining the impact of the number of adaptive rounds.

For the classical equivalence testing problem, in this paper, we make significant strides toward achieving optimal (non-adaptive) query complexity of $O(\log n)$ by allowing only one round of adaptivity.
\paragraph*{Comarison of our algorithm to $\tilde{O}(\log^{12} n)$-query algorithm by~\cite{kamath2019anaconda}:} ~\cite{kamath2019anaconda} provided a non-adaptive tester for the equivalence testing problem, which requires $\Tilde{O}(\log^{12} n)$ queries. Let us briefly describe their algorithm and compare it with ours.  They construct $\Tilde{O}(\log^6 n)$-many sets of varying sizes. Subsequently, they compare the empirical conditional distribution over these subsets for both distributions, by performing $\Tilde{O}(\log^6 n)$ conditional sampling queries on each subsets. If these empirical distributions exhibit significant differences, their algorithm returns reject. On the other hand, if the empirical distributions are close for all such subsets, the algorithm returns accept. 

Despite the similarity in the construction of sets of varying sizes, our algorithm differs in both description as well as analysis.
In our approach, a key component is the role of tuple $(i, S)$ as a certificate that two distributions are far, whereas their algorithm solely considers subsets $S$. Rather than arguing that the empirical distribution over these subsets will differ significantly (if the distributions are far apart), we argue that certain detectable properties will differ with respect to the tuple $(i, S)$. 
This additional dimension allows us to reduce the number of queries significantly. However, it is important to note that our algorithm involves a single round of adaptivity. As a result, our algorithm is incomparable to theirs in this regard.

\section{An Efficient One-Round Adaptive Algorithm}
\label{sec:main-algorithm}
Our main contribution is a one-round adaptive tester for the equivalence testing problem in the {\COND} model that makes at most $\Tilde{O}(\log n)$ queries.
\begin{theorem}
    There exists an algorithm which, given  $\COND$ access to two distributions $\mcP$ and $\mcQ$ on $[n]$ and a parameter $\epsilon >0$,  makes at most $O\left(\frac{\log n (\log \log n) \log 1/\epsilon}{\epsilon^9}\right)$  queries to the oracle and is one-round adaptive, and decides whether $\mcP=\mcQ$ or $d_{TV}(\mcP,\mcQ) \ge \epsilon$ with probability at least $2/3$.
    
  %  and is one-round adaptive and makes at most $O(\log n (\log \log n) \log 1/\epsilon/\epsilon^8)$ conditional queries and with probability at least $2/3,$ it returns {\Accept} if $\mcP = \mcQ$ and returns {\Reject} if $d_{TV}(\mcP,\mcQ) \ge \epsilon.$
\end{theorem}

In the remaining part of this paper, we will prove the above theorem. We start with a high-level idea behind our algorithm, and then we provide a formal description of the algorithm with a detailed analysis.
\subsection{High-Level Overview}

%We now briefly explain the previous our algorithm~\cite{} for equivalence testing. The algorithm first selects  a random power of $2$ between $2$ and $n$. Then it chooses a random set $S \subseteq [n]$ of this size. Now it performs $O(\log^6 n)$ $COND_p(S)$ and $COND_q(S)$ queries and form an empirical distribution $\hat{p}_S$ and $\hat{q}_S$ on $S$. The above procedure is repeated $O(\log^6 n)$ times. If  every time the empirical distributions $\hat{p}_S$ and $\hat{q}_S$ is close in $l_{\infty}$ distance  then the algorithm concludes that distributions $p$ and $q$ are same. Hence the query complexity of their algorithm is $O(\log^{12} n)$
        
%To illustrate the main ideas in our algorithm, let us assume that for any $i,j \in [n]$, the algorithm can determine if $D(i) > D(j)$ or $D(i) \le D(j)$. Note that by $\COND(\{i,j\})$ queries sufficiently number of times, it is possible to determine if $D(i) \ge D(j) +\delta$ or $D(i) \le D(j) - \delta$ for $\delta > 0$.  

The first attempt to design an equivalence tester is to pick a sample, say $i$, from $\mcP$ and compare the probability mass $\mcP(i)$ and $\mcQ(i)$ in $\mcP$ and $\mcQ$ respectively. If $\mcP$ and $\mcQ$ are far (in total variation distance), then with enough probability, the probability mass of $i$ in both distributions is significantly different. However, the issue is that estimating the probability mass of the element $i$ can be very expensive. To bypass this issue, the idea is to sample a subset $S$ of the domain, hoping the following: 
\begin{itemize}
    \item the probability mass of $S$ in $\mcP$ is comparable to that of the mass of $S$ in $\mcQ$, and
    \item  the probability mass of $i$ (in $\mcP$) is similar to the probability mass of $S$.
\end{itemize}
Assuming that all the above two statements hold, one can use conditional sampling (conditioned on $S\cup i$) to compare $\mcP(i)/\mcP(S\cup i)$ and $\mcQ(i)/\mcQ(S\cup i)$. Since, we assumed $\mcP(S)$ is similar to $\mcQ(S)$ so with high enough probability  $\mcP(i)/\mcP(S\cup i)$ and $\mcQ(i)/\mcQ(S\cup i)$ will be different enough. And since we assumed that $\mcP(i)$ is comparable to $\mcP(S)$, one can estimate $\mcP(i)/\mcP(S\cup i)$ and can upper bound $\mcQ(i)/\mcQ(S\cup i)$ using only a few samples, which should be sufficient to distinguish $\mcP$ from $\mcQ$ if the the two distributions are far. But the issue is how to take care of the two assumptions. 

Firstly, for the second assumption, since we don't know the quantity $\mcP(i)$ beforehand, trying to pick a $S$ with similar probability seems unrealistic. For this, we pick a collection of sets, $S_1, \dots, S_{\log n}$, where the set $S_k$ is obtained by picking each element of the domain with probability $1/2^k$. This ensures that the expected value of $\mcP(S_k)$ is $1/2^k$. So, irrespective of what the value of $\mcP(i)$ is, there exists (with high probability) a $S^*$ such that $\mcP(S^{*})$ is comparable to $\mcP(i)$.  
Thus, the hope is to go over all the sets and estimate the $\mcP(i)/\mcP(S_k\cup i)$ and  $\mcQ(i)/\mcQ(S_k\cup i)$ for all the $\log n$ sets, as long as the ratios are within a particular range. Assuming that $\mcP(S)$ and $\mcQ(S)$ is comparable and $\mcP$ and $\mcQ$ are far, the value of  $\mcP(i)/\mcP(S^{*}\cup i)$ and $\mcQ(i)/\mcQ(S^{*}\cup i)$ will be different enough. 

For the above argument to go through, we need the other assumption that the probability mass of $S$ in $\mcP$ is comparable to that of the mass of $S$ in $\mcQ$. Since the sets $S_k$ are obtained by independently drawing elements from the domain, one expects the assumption to hold. While the expected weight of $S$ according to $\mcP$ and $\mcQ$ will be the same, we need to prove concentration. The concentration is hard to achieve in this case. In other words, when the random set $S_k$ is drawn the expected value of $\mcP(S_k) = \mcQ(S_k) = 1/2^k$, and either of the two cases can happen: 
\begin{itemize}
    \item (Case 1) The value of the random variable $\mcP(S_k) - \mcQ(S_k)$ is concentrated around $0$, or
    \item (Case 2) There is large ``tail probability," because of which concentration is not possible. (The notion of tail probability is formalized in Section~\ref{sec:technical-analysis}).
\end{itemize}
If Case 1 holds, that is, the value of the random variables $\mcP(S_k)$ and  $\mcQ(S_k)$ are concentrated around the expectation then the argument of the previous para goes through and we will be able to distinguish $\mcP$ from $\mcQ$ by estimating $\mcP(i)/\mcP(S^{*}\cup i)$ and $\mcQ(i)/\mcQ(S^{*}\cup i)$. 

On the other hand, if Case 2 holds, then it means that the tail probability is high, and if this happens, it means $\mcP$ and $\mcQ$ are far. This case can be caught by estimating the tail probability. This is what is done in our algorithm $\EstTail$. 

So our main tester $\EquivTester$ first picks a number of samples according to $\mcP$ and then constructs $\tilde{O}(\log n)$ sets $S_k$. For each set $S_k$ and for each sample $i$ it estimates the difference between $\mcP(i)/\mcP(S_{k}\cup i)$ and $\mcQ(i)/\mcQ(S_{k}\cup i)$ and also the tail probability (using $\EstTail$). If either of the two estimates is large, the algorithm rejects $\mcP$ and $\mcQ$, and if the algorithm does not reject in all the iterations, then the algorithm accepts. 

Note that the power of conditional samples is used in estimating the values of $\mcP(i)/\mcP(S_{k}\cup i)$ and $\mcQ(i)/\mcQ(S_{k}\cup i)$ and also for estimating the tail probabilities. Regarding the amount of adaptiveness used in the algorithm $\EquivTester$, we observe that once the sets $S_k$'s are fixed, the rest of the samples (conditional samples) can be drawn in parallel.

\subsection{Algorithm Description}
Our algorithm, \EquivTester{}, takes as input, two distributions $\mcP$ and $\mcQ$, and a parameter $\epsilon > 0$. It returns \Accept{} if $\mcP = \mcQ$ and \Reject{} if their total variation distance $d_{TV}(\mcP,\mcQ)$ is greater than $\epsilon$, both with at least $2/3$ probability.

\EquivTester{} samples $O(1/\epsilon)$ points from $\mcP$, with the set of all such points denoted by $E$ (line~\ref{line:def-E}). It then constructs subset $S_t$ for each $t$ in $\{1,1/2,1/4,\dots,1/n\}$, such that each element from $[n]$ is included in $S_t$ with probability $t$ (lines~\ref{line:alg-tester-for-prob-begin}--\ref{line:alg-tester-for-prob-end}). The algorithm then employs two subroutines, \EstProb{} and \EstTail{}, for each tuple $(i,S)$ (where $i \in E$ and  $S = S_t$ for some $t$) (lines~\ref{line:start-for}--\ref{line:end-for}). We invoke {\EstProb} to estimate corresponding conditional probabilities $\mcP(i)/\mcP(i \cup S)$ and $\mcQ(i)/\mcQ(i \cup S)$ and return Reject if the difference between conditional probabilities is far (lines~\ref{line:compare-prob-start}--\ref{line:compare-prob-end}). If the difference is not far, we invoke {\EstTail} (to estimate the tail probability {\TP}, formally defined in Section~\ref{sec:technical-analysis}) and again, we reject if the difference between the tail probabilities of $(i,S)$ for  $\mcP$ and $\mcQ$ is far (lines~\ref{line:compare-tail-start}--\ref{line:compare-tail-end}). Finally, if for all tuples $(i,S)$, all these estimates are close, {\EquivTester} returns \Accept (line~\ref{line:Accept}).

\paragraph*{Query complexity:} We now establish an upper bound on the number of calls to the $\COND$ oracle by the {\EquivTester} algorithm. Note that each invocation of $\EstProb$ results in $m = O\left(\frac{(\log \log n) \log 1/\epsilon}{\epsilon^8}\right)$ calls to the $\COND$ oracle, and each invocation of $\EstTail$ leads to $mb = 40000m$ calls to the $\COND$ oracle. Given that {\EquivTester} invokes both $\EstProb$ and $\EstTail$ at most $|E \times \mathcal{S}| \le 20 \log n/\epsilon$ times, the total number of calls made to the $\COND$ oracle is at most $O\left(\frac{\log n (\log \log n) \log 1/\epsilon}{\epsilon^9}\right)$. 

\paragraph*{Making {\EquivTester} one-round adaptive:} For the sake of improved presentation of {\EquivTester}, we have opted not to group together the conditional queries that can be made simultaneously. We now modify {\EquivTester} by re-arranging the order of the conditional queries and making it a one-round adaptive algorithm.   First, we note that the construction of $\mathcal{S} = \{S_t: t \in \{1, 1/2, \dots, 1/n\}\}$ does not require any call to the  $\COND$ oracle.  %Additionally, it's worth emphasizing that all conditional queries executed by {\EquivTester} take place at the following lines: line~\ref{line:def-E} of {\EquivTester}, line~\ref{line:EstProb-start} of {\EstProb},  line~\ref{line:EStTatil-start} of {\EstTail} and line~\ref{line:third-round} of {\EstTail}.

To convert {\EquivTester} into a one-round algorithm, all conditional queries executed in line~\ref{line:def-E} of {\EquivTester} and line~\ref{line:EStTatil-start} of {\EstTail} can be made simultaneously. This is possible since these queries are either of the form $\COND_{\mcP}([n])$ or $\COND_{\mcP}(S)$ and $\COND_{\mcQ}(S)$ for some $S \in \mathcal{S}$ and as noted before, the set $\mathcal{S}$ can be constructed beforehand.

Then, we can make all the remaining conditional queries simultaneously.  These queries are of the form: (i) $\COND_{\mcP}(i \cup S)$ and $\COND_{\mcQ}(i \cup S)$ (line~\ref{line:EstProb-start} of {\EstProb}) where $i \sim \COND_{\mcP}([n])$ is the outcome of query made in the previous round and $S \in \mathcal{S}$ is available beforehand, (ii)  $\COND_{\mcP}(i \cup j)$ and $\COND_{\mcQ}(i \cup j)$ (line \ref{line:third-round} of {\EstTail}) where again  $i \sim \COND_{\mcP}([n])$ and $j \sim \COND_{\mcD}(S)$ (for some $S \in \mathcal{S}$ and $\mcD \in \{\mcP,\mcQ\}$) are the outcomes from the queries in the first round.

\subsection{Technical Analysis}\label{sec:technical-analysis}

Before we formally present the analysis of the correctness and complexity of {\EquivTester}, we define the tail probability, $\TP{}$. Given a distribution $\mcD$, a tuple $(i, S)$ where $i \in [n]$ and $S \subseteq [n]$, and parameters $\beta \in (0,1)$ and $b \in \mathbb{Z}^+$, the tail probability $\TP(\mcD, i, S, \beta, b)$ is defined as follows:

It is the probability that a random sample $j \sim \COND_{\mcD}(S)$ will occur no more than $\frac{1}{2} + \beta$ times in $b$ independent queries of $\COND_{\mcD}({i, j})$. Formally, it can be expressed as:
\begin{align*}
 &   \TP(\mcD,i,S,\beta,b) := \nonumber\\
 &  \Pr_{j \sim \COND_{\mcD}(S)} \left[\bin\left(b,\frac{\mcD(j)}{\mcD(j)+\mcD(i)}\right) \le (\frac{1}{2}+\beta)b\right] \nonumber\\
 & = \sum_{j \in S} \frac{\mcD(j)}{\mcD(S)}  \Pr\left[\bin\left(b,\frac{\mcD(j)}{\mcD(j)+\mcD(i)}\right) \le (\frac{1}{2}+\beta)b\right].  %\Pr[\hat{D}(j, \{i\},B) \le \frac{1}{2}+\beta]
\end{align*}

We now analyze our algorithms. The subroutine $\EstProb(\mcD,i,S,m)$ takes as input a distribution $\mcD$, $i \in [n]$, $S \subseteq [n]$ and a parameter $m \in \mathbb{Z}^+$. It uses a straightforward estimator to return $\pm \gamma$ estimate of $\frac{\mcD(i)}{\mcD(i \cup S)}$ with probability at least $1 - e^{-\gamma^2 m}$.

\begin{lemma}
    \label{lem:estimatorD(i)/D(i)+D(S)}
    For an arbitrary $i \in [n]$, $S \subseteq [n]$, $\gamma \in  (0,1)$ and $m \ge 1$ and distribution $\mcD$, the $\EstProb(\mcD,i,S,m)$ returns $\pm \gamma$ estimate of $\frac{\mcD(i)}{\mcD(i \cup S)}$ with probability at least $1 - e^{-\gamma^2 m}$.
\end{lemma}

%   We now formally show that  \EstProb{} and \EstTail{}  estimates  $D(i|\{i\} \cup S)$ and the tail probability $\TP$ respectively.

%   \todo{say why it is one round adaptive}
\begin{algorithm}[H]
\caption{$\EquivTester(\mcP,\mcQ,\epsilon)$}
\hspace*{\algorithmicindent} \textbf{Input}: A pair of distribution $\mcP,\mcQ$ on $[n]$, $\epsilon >0$ \\
 \hspace*{\algorithmicindent}\textbf{Output}: \Accept{} with prob. $2/3$ if $\mcP = \mcQ$, \Reject{} with prob. $2/3$ if $d_{TV}(\mcP,\mcQ) \ge \epsilon$.
\begin{algorithmic}[1]
        \State $\gamma \gets \frac{\epsilon^4}{L}$ where $L \gets 10^{15}$ is a large constant.
        \State $m \gets 100  (\log \log n) \log \frac{1}{\epsilon}/\gamma^2$.
        
        \State $\beta = 0.05, b = 100/\beta^2$.
        \State Sample $20/\epsilon$ points from $\mcP$. Let  $E$ be the set of such points.\label{line:def-E}
   %     \State $E \gets 10/\epsilon$
       % \State  Sample $n_1=O(\frac{1}{\epsilon})$ points $i_1,\dots,i_{n_1}$ from  the distribution $P$.
    %    \For {$i \in [E]$}
    %        \State Sample $s_i \sim D_1$.
     %   \EndFor
         %   \State $CompareMass(i,[n]\setminus i)$
         %   \State $CompareTail(i,[n]\setminus i)$
        \For{$t \in \{1, \frac{1}{2}, \frac{1}{4},\dots,\frac{1}{n}\}$ }\label{line:alg-tester-for-prob-begin}
                    \State Construct set $S_t$ by picking each element of $[n]$ independently with probability $t$.
        \EndFor\label{line:alg-tester-for-prob-end}
        \State Let $\mathcal{S} = \{S_t| t \in \{1, \frac{1}{2}, \frac{1}{4},\dots,\frac{1}{n}\}\}.$
%        \Comment{If any pair $(i,S)$ is a distinguisher then the below for loop will \Reject{} whp.}
        \For{all tuple $(i,S) \in E \times \mathcal{S}$ }\label{line:start-for}
                 \State $\epP = \EstProb(\mcP,i,S,m).$\label{line:compare-prob-start}
                 \State $\epQ = \EstProb(\mcQ,i,S,m).$
                 \If {$|\epP - \epQ| > 2\gamma$} 
                     \State \Return \Reject \label{line:compare-prob-end}
                 \EndIf
                 \State $\etP = \EstTail(\mcP,i,S,\beta,b,m).$\label{line:compare-tail-start}
                 \State $\etQ = \EstTail(\mcQ,i,S,\beta,b,m).$
                 \If{$|\etP - \etQ| > 2 \gamma$}
                    \State \Return \Reject \label{line:compare-tail-end}
                 \EndIf
             %       \State $m \gets m+1$
           %     \EndWhile
        \EndFor \label{line:end-for}
        \State \Return \Accept\label{line:Accept}.
  \end{algorithmic}
\end{algorithm}

%We call the line `If ($|EstProb(D_1,i,S,M) - EstProb(D_2,i,S,M)| > 2 \delta$) reject and exit' as $CompareProb(i,S)$ and  line `If ($|EstTail(D_1,i,S,\beta,B,N) - EstTail(D_2,i,S,\beta,B,N)| > 2 \gamma$) reject and exit' as $CompareTail(i,S)$.

\begin{algorithm}
\caption{$\EstProb(\mcD,i,S,m)$}
\hspace*{\algorithmicindent} \textbf{Input}: distribution $\mcD$ on $[n]$, $i \in [n]$, $S \subseteq [n]$, parameter $m \ge 1$ \\
 \hspace*{\algorithmicindent} \textbf{Output}: estimate  of $\frac{\mcD(i)}{\mcD(i \cup S)}$
\begin{algorithmic}[1]
   %     \State $\gamma = \frac{\epsilon^{5}}{1000 ( \log  \log n)^4}, M = \frac{( \log \log n)^2 + \log \frac{1}{\epsilon}}{\gamma^2} $.
        \State Sample $j_1,\dots,j_m \sim \COND_{\mcD}(i \cup S)$.\label{line:EstProb-start}
       % \State Let $C = \sum_{k \in [m]} \mathbbm{1}_{X_k = i}$.
        \State  return $ \frac{\sum_{k \in [m]} \mathbbm{1}_{j_k = i}}{m}$
  %      \State return 
%         \State Perform $COND_P(S \cup i)$ queries  $n_3   = \frac{(\log \log \log n)^2 + \log \frac{1}{\epsilon}}{\delta_3^2}$ times.
%         \State Let $\hat{E}_P(i,S)=\frac{\text{\# times}\thinspace i \thinspace \text{appears} }{n_3}$.
%         \State Perform $COND_Q(S \cup i)$ queries  $n_3$ times.
%         \State Let $\hat{E}_Q(i,S)=\frac{\text{\# times} \thinspace i \thinspace \text{appears} }{n_3}$.
 %        \If{$|\hat{P}_1(i, S,n_3) - \hat{P}_2(i,S,n_3)|>  2 \delta_3$}
%            \State reject and exit.
 %        \EndIf
\end{algorithmic}   
\end{algorithm}

 \begin{algorithm}
\caption{$\EstTail(\mcD,i,S,\beta,b,m)$}
\hspace*{\algorithmicindent} \textbf{Input}: distribution $\mcD$ on $[n]$, $i \in [n]$, $S \subseteq [n]$, parameters $\beta >0$, $b,m \ge 1$ \\
 \hspace*{\algorithmicindent} \textbf{Output}: estimate  of $\TP(\mcD,i,S,\beta,b)$
\begin{algorithmic}[1]
   %     \State $\gamma = \frac{\epsilon^{5}}{1000 ( \log  \log n)^4}, M = \frac{( \log \log n)^2 + \log \frac{1}{\epsilon}}{\gamma^2} $.
        \State Sample $j_1,\dots,j_m \sim \COND_{\mcD}(S)$.\label{line:EStTatil-start}
        \For{$k \in [m]$}
            \State Sample $y_1,\dots,y_b \sim \COND_{\mcD}(\{j_k, i\})$.\label{line:third-round}
            \If{$\frac{\sum_{\ell \in [b]}\mathbbm{1}_{y_\ell = j_k}}{b} \le \frac{1}{2} + \beta$}
                \State $Z_k = 1$.
            \Else
                \State $Z_k=0$
            \EndIf
        \EndFor
        \State $Z = \frac{\sum_{k \in [m]}Z_k}{m}$.
        \State return $Z$.
            
    %    \State $\hat{D}(i,S) = \frac{C_i}{M}$
%         \State Perform $COND_P(S \cup i)$ queries  $n_3   = \frac{(\log \log \log n)^2 + \log \frac{1}{\epsilon}}{\delta_3^2}$ times.
%         \State Let $\hat{E}_P(i,S)=\frac{\text{\# times}\thinspace i \thinspace \text{appears} }{n_3}$.
%         \State Perform $COND_Q(S \cup i)$ queries  $n_3$ times.
%         \State Let $\hat{E}_Q(i,S)=\frac{\text{\# times} \thinspace i \thinspace \text{appears} }{n_3}$.
 %        \If{$|\hat{P}_1(i, S,n_3) - \hat{P}_2(i,S,n_3)|>  2 \delta_3$}
%            \State reject and exit.
 %        \EndIf
\end{algorithmic}   
\end{algorithm}

\iffalse
\begin{proof}
Note that for any $k \in [m]$, $\mathbbm{1}_{j_k =i}$ is a Bernoulli random variable such that $\Pr[\mathbbm{1}_{j_k =i} =1] = \Pr[j_k = i] = \frac{\mcD(i)}{\mcD(i \cup S)}$. Also, note that  $\mathbbm{1}_{j_k =i}$  are iid for all $k\in [m]$. By additive Chernoff bound (Lemma~\ref{lem:additiveChernoffBound}), the value $\mathsf{ep} = \EstProb(\mcD,i,S,m)$ returned by  the estimator  satisfies:
\begin{align*}
%\label{eq:Chernoff-CompareMass}
\Pr\left[\left|\mathsf{ep} - \frac{\mcD(i)}{\mcD(i \cup S)} \right| > \gamma\right] \le e^{-\gamma^2 m}.
\end{align*}
    
\end{proof}

\fi

The subroutine $\EstTail(\mcD,i,S,\beta,b,m)$ takes as input a distribution $\mcD$, $i \in [n]$, $S \subseteq [n]$ and  parameters $\beta \in (0,1), b,m \in \mathbb{Z}^+$. It returns, in a straightforward way, $\pm \gamma$ estimate of $\TP(\mcD,i,S,\beta,b,m)$ with high probability.
\begin{lemma}
    \label{lem:estimatorTailProb}
  For any  distribution $\mcD$ on $[n]$, $i \in [n]$, $S \subseteq [n]$, parameters $\beta \in (0,1)$, $b,m \in \mathbb{Z}^+$,  the $\EstTail(\mcD,i,S,\beta,b,m)$ returns $\pm \gamma$ estimate of $\TP(\mcD,i,S,\beta,b)$ with probability at least $1 - e^{-\gamma^2 m}$.
\end{lemma}

\iffalse
\begin{proof}
For any $k \in [m]$, we have
\begin{align*}
 &   \Pr[Z_k = 1] \\
  &  = \sum_{j \in S} \Pr[j_k = j] \Pr\left[\sum_{\ell \in [b]}\mathbbm{1}_{y_\ell = j} \le \left(1/2+\beta \right)b\right]\\
  &   =  \sum_{j \in S} \frac{\mcD(j)}{\mcD(S)} \Pr\left[\bin\left(b,\frac{\mcD(j)}{\mcD(i)+\mcD(j)}\right) \le (1/2+\beta)b\right]\\
  &    = \TP(\mcD,i,S,\beta,b).
\end{align*}
Therefore, by additive Chernoff bound (Lemma~\ref{lem:additiveChernoffBound}), the value $\mathsf{et} = \EstTail \left(\mcD,i,S,\beta,b,m \right)$ returned by the algorithm satisfies
\begin{align*}
%\label{eq:Chernoff-CompareProb}
\Pr[\left|\mathsf{et} - \TP \left(\mcD,i,S,\beta,b \right) \right| > \gamma] \le e^{-\gamma^2 m}.
\end{align*}

 %   Fix any $X_k$ for $k \in [N]$. We have $\Pr[X_k = j] = \frac{D(j)}{D(S)}$ for any $j \in S$. Also for any $\ell \in [B]$, $\Pr[1_{Y_\ell = X_k}] =\Pr[Y_\ell = X_k] = \frac{D(X_k)}{D(i)+D(X_k)}.$  Thus  $\sum_{\ell \in [B]}1_{Y_\ell = X_k}$ is distributed as binomial random variable $Bin(B,\frac{D(X_k)}{D(i)+D(X_k)})$ and the random variable $Z_k$ is a Bernoulli random variable with success probability $\Pr[Bin(B,\frac{D(X_k)}{D(i)+D(X_k)}) \le (1/2+\beta)B]$.
\end{proof}

\fi

The proofs of both Lemma~\ref{lem:estimatorD(i)/D(i)+D(S)} and Lemma~\ref{lem:estimatorTailProb} are by a standard application of additive Chernoff bound (Lemma~\ref{lem:additiveChernoffBound}), and deferred to the supplementary materials.

%Since the value of the parameters $\beta,b,m$ are fixed throughout the paper, for brevity, we will omit them in the 
For brevity, from now on, we use $\EstProb(\mcD,i,S)$ for $\EstProb(\mcD,i,S,m)$, $\EstTail(\mcD,i,S)$ for $\EstTail(\mcD,i,S,\beta,b,m)$ and $\TP(\mcD,i,S)$ for $\TP(\mcD,i,S,\beta,b,m)$. 
%We can do so because the parameters $m$, $\beta$, and $b$ are implicitly included as input variables and thus need not be explicitly stated.%This is because it is always understood that these parameters are part of inputs.
%In other words, we will omit the parameters $\beta,b,m$ as 

We now prove the first part of our main theorem, i.e., if $\mcP = \mcQ$, then \EquivTester{}  returns \Accept{} with high probability.

\begin{lemma}\label{lem:accept-same}
If $\mcP = \mcQ$ then the algorithm returns \Accept{} with probability at least $1-o(1)$.
\end{lemma}
\begin{proof}
For each $(i,S) \in E \times \mathcal{S},$  let $\bad_1(\mcP,i,S)$ be the event that
\begin{align*}
    \left|\EstProb(\mcP,i,S) - \frac{\mcP(i)}{\mcP(i \cup S)}\right| \ge \gamma
\end{align*}
and $\bad_2(\mcP,i,S)$ be the event that
\begin{align*}
     \left|\EstTail(\mcP,i,S) - 
     \TP(\mcP,i,S)\right| \ge \gamma.
\end{align*}
Similarly, we define the events $\bad_1(\mcQ,i,S)$ and $\bad_2(\mcQ,i,S)$. Now, consider the event $\bad:= \bigcup_{(i,S) \in E \times \mathcal{S}} (\bad_1(\mcP,i,S) \cup \bad_2(\mcP,i,S) \\$
$\cup \bad_1(\mcQ,i,S) \cup \bad_2(\mcQ,i,S))$.

From Lemma~\ref{lem:estimatorD(i)/D(i)+D(S)} (substituting $m = 100  (\log \log n) \log \frac{1}{\epsilon}/\gamma^2$),  we have $\Pr\left[\bad_1(\mcP,i,S)\right] \leq e^{-\gamma^2 m}$ and $\Pr\left[\bad_1(\mcQ,i,S)\right] \leq e^{-\gamma^2 m}$. Similarly, from Lemma~\ref{lem:estimatorTailProb}, we have  $\Pr\left[\bad_2(\mcP,i,S)\right] \leq e^{-\gamma^2 m} $ and $\Pr\left[\bad_2(\mcQ,i,S)\right] \leq e^{-\gamma^2 m}$.

Since the total number of  $(i,S)$ tuples considered by the algorithm is at most $10 \log n \cdot \epsilon^{-1}$, by a union bound, we have $\Pr[\bad] \le  10 \log n \cdot \epsilon^{-1} \cdot 4 \cdot e^{-\gamma^2 m} \le  1/(\log n)^{98}$ (since $e^{-\gamma^2 m}  \le \epsilon / ( \log n)^{100}$). Further, if the event $\bad$ does not happen, then for all $(i,S) \in E \times \mathcal{S},$ we have $ |\EstTail(\mcP,i,S) -  \EstTail(\mcQ,i,S) |\le 2 \gamma$ and $ |\EstProb(\mcP,i,S) -   \EstProb(\mcQ,i,S) | \le 2 \gamma$. Hence, with probability at least  $1-o(1)$, \EquivTester{} will return \Accept.
\end{proof}

We now proceed towards showing the second part of our main theorem, if $d_{TV}(\mcP,\mcQ) \ge \epsilon$, then {\EquivTester} returns  \Reject{} with high probability.
\begin{lemma}\label{lem:reject-different}
If $d_{TV}(\mcP,\mcQ) \ge \epsilon$, then {\EquivTester} returns {\Reject} with probability at least $2/3$.
\end{lemma}

\begin{proof}
We start with the notion of a tuple $(i,S) \in E \times \mathcal{S}$ being a \emph{distinguisher} for $\mcP$ and $\mcQ,$ which will prove to be a sufficient condition for {\EquivTester} to return {\Reject} with high probability.
\begin{definition}
    \label{def:distinguisher}
   A tuple $(i,S) \in E \times \mathcal{S}$  is called a \emph{distinguisher} for $\mcP$ and $\mcQ$ if either of the following two conditions hold true:
   \begin{enumerate}
       \item $\left| \frac{\mcP(i)}{\mcP(i \cup S)} - \frac{\mcQ(i)}{\mcQ(i \cup S)} \right| > 4 \gamma$,  \label{item:dist-prob}
       \item $|\TP(\mcP,i,S) - \TP(\mcQ,i,S)| > 4\gamma.$ \label{item:dist-tail}
   \end{enumerate}
\end{definition}

To complete the proof, we rely on the following three lemmas, whose proofs we will provide later.
\begin{lemma}\label{lem:if-distinguisher-then-reject}
If $(i,S) \in E \times \mathcal{S}$ is a distinguisher for $\mcP$ and $\mcQ$, then {\EquivTester} returns {\Reject} with  probability at least $1-4/(\log n)^{100}.$ 
\end{lemma}

% Given Lemma~\ref{lem:if-distinguisher-then-reject}, it suffices for us to show that with high probability, there exists a tuple $(i,S) \in E \times \mathcal{S}$ that is a distinguisher for $\mcP$ and $\mcQ.$  We will first show that (with high probability) there will exist a point (which we will call $i^*$ throughout the paper) in $E$ with some desirable properties (Lemma~\ref{lem:qilesspi}). The point $i^*$ will be the first component of the distinguisher, i.e., subsequently, we will show a set $S \in \mathcal{S}$ such that the tuple $(i^*,S)$ is  a distinguisher.

%Now assume that $d_{TV}(D_1,D_2) \ge \epsilon$. We need to show that the algorithm rejects with high probability.
\begin{lemma}
\label{lem:qilesspi}
Let $c = 1000$. If $d_{TV}(\mcP,\mcQ) \ge \epsilon$, then with probability at least $1-e^{-6}$, there exists a non-empty set  $\mathcal{T} \subseteq E$, such that for all $i^* \in \mathcal{T}$, we have 

% (i) $   \sum_{j:\mcP(j) \le \mcP(i^*)}\mcP(j) \ge 3\epsilon/10$, (ii) $\mcP(i^*) \ge (1+ \epsilon/4) \mcQ(i^*)$, and (iii) $\mcP(i^*) \ge \frac{\epsilon^3}{c^2 n}$.
\begin{enumerate}
    \item \label{item:taili^*bound}
 $   \sum_{j:\mcP(j) \le \mcP(i^*)}\mcP(j) \ge 3\epsilon/10$,

%\begin{align}\label{eq:taili^*bound}
%    \sum_{j:D_1(j) \le D_1(i^*)}D_1(j) \ge \epsilon/8
%\end{align}
\item \label{item:D1largeD2}
    $\mcP(i^*) \ge (1+ \epsilon/4) \mcQ(i^*)$,
%\begin{align} \label{eq:D1largeD2}
%    D_1(i^*) \ge (1+ \epsilon/4) D_2(i^*)
%\end{align}

\item  \label{item:i^*nottoosmall}
    $\mcP(i^*) \ge \frac{\epsilon^3}{c^2 n}$.
 \end{enumerate}
% If we sample $i \sim D_1$ then with probability at least $\epsilon/8$, we have $D_1(i) \ge (1+ \epsilon/4) D_2(i),$  $Tail_D(i,[n],\infty) =  \sum_{j:D_1(j) \le D_1(i)}D_1(j) \ge \epsilon/8$ and $D_1(i) \ge \frac{\epsilon^3}{(\log \log  n)^3 n}$.
\end{lemma}
%Let $S^* = S_{t^*}$, i.e., $S^*$ is the set constructed by picking each element in $[n]$ with probability $t^*$.

Let us now consider the subset $\mathcal{T} \subseteq E$ from the above lemma.

%We will show that with high probability, for every   $i^* \in \mathcal{T},$  there exists a $S^* \in \mathcal{S}$ such that  $(i^*,S^*)$ will be a distinguisher for  $\mcP$ and $\mcQ$.
\begin{lemma}
    \label{lem:eitherProborTail}
For every $i^* \in \mathcal{T}$, with probability at least $4/5$, there exists a $S^* \in \mathcal{S}$, such that    $(i^*,S^*)$ is a distinguisher for $\mcP$ and $\mcQ$.
\end{lemma} 

We are now ready to finish the proof of Lemma~\ref{lem:reject-different}. It now directly follows from Lemma~\ref{lem:qilesspi}, Lemma~\ref{lem:eitherProborTail}, and Lemma~\ref{lem:if-distinguisher-then-reject}, that if
$d_{TV}(\mcP,\mcQ) \ge \epsilon$, then {\EquivTester} does not return {\Reject} with probability  at most $ e^{-6} + 1/5 + 4/(\log n)^{100} < 1/3$. %In other words, {\EquivTester} returns {\Reject} with probability at least $2/3.$
%, wherein:
% \begin{enumerate}
%  \item $e^{-6}$ accounts for the fact that $\mathcal{T}$ may be empty (Lemma~\ref{lem:qilesspi}), 
%  \item $4/(\log n)^{100}$ accounts for the fact that {\EquivTester} may not return {\Reject} for a distinguisher (Lemma~\ref{lem:if-distinguisher-then-reject}), and
%  \item $1/5$ accounts for the fact that $(i^*,S^*)$ may not be distinguisher (Lemma~\ref{lem:eitherProborTail}).  \todo{Still needs revision}
% \end{enumerate}
% (1) $e^{-6}$ accounts for the fact that $\mathcal{T}$ may be empty (Lemma~\ref{lem:qilesspi}), (2) $4/(\log n)^{100}$ accounts for the fact that {\EquivTester} may not return {\Reject} for a distinguisher (Lemma~\ref{lem:if-distinguisher-then-reject}), and (3) $1/5$ accounts for the fact that $(i^*,S^*)$ may not be distinguisher (Lemma~\ref{lem:eitherProborTail}). 
\end{proof}

%\subsection{Proof of Lemma~\ref{lem:reject-different}}

% \begin{lemma}\label{lem:if-distinguisher-then-reject}
% If $(i,S) \in E \times \mathcal{S}$ is a distinguisher for $\mcP$ and $\mcQ$ then    {\EquivTester} returns {\Reject} with  probability at least $1-4/(\log n)^{100}.$ \todo{Change the font}
% \end{lemma}
\paragraph*{Proof of Lemma~\ref{lem:if-distinguisher-then-reject}}
\begin{proof}
Analogous to the proof of Lemma~\ref{lem:accept-same}, we now define the events $\bad_1(\mcP), \bad_2(\mcP)$, and so on.
    For any distribution $\mcD \in \{\mcP,\mcQ\},$    let $\bad_1(\mcD)$ be the event that
\begin{align*}
    \left|\EstProb(\mcD,i,S) - \frac{\mcD(i)}{\mcD(i \cup S)}\right| \ge \gamma
\end{align*}
and $\bad_2(\mcD)$ be the event that
\begin{align*}
     \left|\EstTail(\mcD,i,S) - 
     \TP(\mcD,i,S)\right| \ge \gamma.
\end{align*}
 Let $\bad:=   \cup_{\mcD \in \{\mcP,\mcQ\}} \cup_{j\in \{1,2\}} \bad_j(\mcD).$

Since $m = 100  (\log \log n) \log \frac{1}{\epsilon}/\gamma^2$, from Lemma~\ref{lem:estimatorD(i)/D(i)+D(S)} and Lemma~\ref{lem:estimatorTailProb},  we have for $\mcD \in \{\mcP,\mcQ\}$, both $\Pr\left[\bad_1(\mcD)\right]$ and $\Pr\left[\bad_2(\mcD)\right]$ are at most $e^{-\gamma^2 m}  \le \epsilon/( \log n)^{100}$. Thus by a union bound, the event $\bad$ happens with probability at most $4\epsilon/(\log n)^{100}$.

  Since $(i,S)$ is fixed in the context of this lemma,  for brevity, we use $\epP$ for $\EstProb(\mcP,i,S)$ and $\epQ$ for $\EstProb(\mcQ,i,S)$ for the remaining parts of the proof of this lemma. Similarly, we use $\etP$ for $\EstTail(\mcP,i,S)$ and $\etQ$ for $\EstTail(\mcQ,i,S)$.

From now on, assume  $\bad$ does not happen. Since $(i,S)$ is a distinguisher, either the item (\ref{item:dist-prob}) or the item (\ref{item:dist-tail}) in the Definition~\ref{def:distinguisher} holds.   If the item (\ref{item:dist-prob}) holds, then by the triangle inequality, we have 
\begin{align*}
& |\epP- \epQ|  \ge  \\
 &  \left|\frac{\mcP(i)}{\mcP(i \cup S)} -  \frac{\mcQ(i)}{\mcQ(i \cup S)}\right| -  \left|\epP - \frac{\mcP(i)}{\mcP(i \cup S)}\right|  \\
& - \left| \epQ - \frac{\mcQ(i)}{\mcQ(i \cup S)}\right|
> 2\gamma
\end{align*}
and thus {\EquivTester} returns {\Reject}.

Now, if the item (\ref{item:dist-tail}) holds, then again, by the triangle inequality, we have 
\begin{align*}
  |\etP - & \etQ| \ge  \\
   & |\TP(\mcP,i,S) - \TP(\mcQ,i,S)|\\  & -   |\etP -   \TP(\mcP,i,S)| -|\etQ - \TP(\mcQ,i,S)| > 2\gamma
\end{align*}
and thus {\EquivTester} returns {\Reject}.
\end{proof}

\paragraph*{Proof of Lemma~\ref{lem:qilesspi}}
\begin{proof}
    Let $A = \{i \in [n]: \mcP(i) > \mcQ(i)\}$. We partition $A$ into $A_1 = \{i \in A: \mcP(i) < \epsilon^3/c^2 n\}$, $A_2 = \{i \in A\setminus A_1: \mcP(i) < (1+ \epsilon/4) \mcQ(i)\}$ and $A_3 = A \setminus \{A_1 \cup A_2\}$. Note that any $i \in A_3$, by definition,  will satisfy $\mcP(i) \ge (1+ \epsilon/4) \mcQ(i)$ and $\mcP(i) \ge \frac{\epsilon^3}{c^2 n}$. We now lower bound  $\mcP(A_3).$
Firstly, %observe that: 
    \begin{align*}
 %&     \epsilon \le 
 %\sum_{i \in A} |\mcP(i)-\mcQ(i)| 
 d_{TV}(\mcP,\mcQ) &=    \sum_{i \in A_1} |\mcP(i)-\mcQ(i)| + \sum_{i \in A_2} |\mcP(i)-\mcQ(i)|\\
 &  \hspace{5em}    + \sum_{i \in A_3} |\mcP(i)-\mcQ(i)| \\
 &      \le  n \cdot \frac{\epsilon^3}{c^2 n} + \sum_{i \in A_2} \mcP(i) \cdot \frac{\epsilon}{4} +\sum_{i \in A_3} \mcP(i)\\
 &      \le \frac{\epsilon^3}{c^2} + \frac{\epsilon}{4} + \sum_{i \in A_3} \mcP(i).
    \end{align*}
    Since  $d_{TV}(\mcP,\mcQ) \ge \epsilon$, we have  $\sum_{i \in A_3} \mcP(i) \ge 3\epsilon/5.$ Let $A_4 = \{i \in A_3: \sum_{j\in A_3:\mcP(j) \le \mcP(i)}\mcP(j) \ge 3\epsilon/10\}$. Observe that every $i^* \in A_4$ satisfies all the items of this lemma. Now, we lower bound $\mcP(A_4)$. 
    
    Note that $\mcP(A_3\setminus A_4) < 3\epsilon/10.$ Therefore, $\mcP(A_4) \ge 3\epsilon/5 - 3\epsilon/10 = 3\epsilon/10.$  Thus, the set $\mathcal{T} := E \cap A_4$  is empty with probability at most $(1-3\epsilon/10)^{20/\epsilon} < e^{-6}$ (since $(1-x)^r \le e^{-xr}$). Therefore, with probability at least $1-e^{-6},$  the set $\mathcal{T}$ is non-empty, and any   $i^* \in \mathcal{T}$ satisfies all the items of this lemma.
  %  Note that     
   % Let $\Bar{i}   = \max_{i \in A_3}\{\mcP(i)| \sum_{j \in A_3, \mcP(j) \le \mcP(i)}\mcP(j) < 3\epsilon/10 \} \in A_3$ be the 
\end{proof}

\subsection*{Proof of Lemma~\ref{lem:eitherProborTail}}
\begin{proof}
%From now on, we assume $i^* \in E$ for which all the items \ref{eq:taili^*bound}, \ref{eq:D1largeD2} and \ref{eq:i^*nottoosmall} of Lemma~\ref{lem:qilesspi} holds. Our distinguisher will be of form $(i^*,S)$ for some $S \in \mathcal{S}$. The goal from now on, is to show there exists such $S \in \mathcal{S}.$

Consider an arbitrary $i^* \in \mathcal{T}.$ Our goal is to show that with high probability, there exists a $S^* \in \mathcal{S}$ such that $(i^*,S^*)$ is a distinguisher. From the definition~\ref{def:distinguisher}, it follows that if $\left|\mcP(i^*) - \mcQ(i^*)\right|  > 4 \gamma$ or $\left|\TP(\mcP,i^*,[n]) -  \TP(\mcQ,i^*,[n])\right| > 4 \gamma$,  then $(i^*,[n])$ is a distinguisher (note that $[n] \in \mathcal{S}$). Therefore, we need to focus only on the following cases:
\begin{align}
    \left|\mcP(i^*) - \mcQ(i^*)\right|  \le 4 \gamma \label{eq:pi/pi+pScloseq}\\
    \left|\TP(\mcP,i^*,[n]) -  \TP(\mcQ,i^*,[n])\right| \le 4 \gamma.\label{eq:zpclosezq}
\end{align}
%To take care of the fact that $i^*$ may not be in $E$, we will add the error probability of {\EquivTester} by $e^{-6}$. 

We now give the value of $t^* \in \{1,1/2,\dots,1/n\}$, in terms of $\mcP(i^*)$ such that the tuple  $(i^*,S_{t^*})$ will be a distinguisher with high probability, where recall  $S_{t^*}$ is a set constructed by picking each element in $[n]$ with probability $t^*.$ For the same, first note that:
\begin{align*}
 &   \mcP(i^*) - \frac{\mcP(i^*)}{1+\epsilon/4}    \le  \mcP(i^*) - \mcQ(i^*) \le 4 \gamma
\end{align*}
where the first inequality is by   Lemma~\ref{lem:qilesspi} and the second by the Eq.~\ref{eq:pi/pi+pScloseq}.  Immediately, we get:
\begin{align}
  &    \mcP(i^*) \le \frac{32 \gamma}{\epsilon} \le \frac{32 \epsilon^3}{L} \label{eq:pismall}.
\end{align}
Let $t' = \frac{c^2 \mcP(i^*)}{\epsilon^3}.$  Since $\mcP(i^*) \le \epsilon^3/c^2 n$, we have $\frac{1}{n} \le t' \le 1$. Therefore, there exists $t^* \in \left\{\frac{1}{n},\frac{2}{n}, \dots,1\right\}$ such that $t' \le t^* < 2t'$, i.e.,
\begin{align}\label{eq:t^*-P(i^*)}
 \frac{c^2 \mcP(i^*)}{\epsilon^3} \le     t^* < 2  \frac{c^2 \mcP(i^*)}{\epsilon^3}.
\end{align}

Let $S^* = S_{t^*}$, i.e., $S^*$ is the set constructed by picking each element in $[n]$ with probability $t^*$. Our goal now is to prove, with probability at least $4/5,$ $(i^*,S^*)$ is a distinguisher for $\mcP$ and $\mcQ.$

% We will show that with high probability, $(i^*,S^*)$ will be a distinguisher for  $\mcP$ and $\mcQ$.
% \begin{lemma}
%     \label{lem:S(t^*)-distinguisher}
%     Let $S^* = S_{t^*}$, i.e., $S^*$ is the set constructed by picking each element in $[n]$ with probability $t^*$. With probability at least $4/5$,  $(i^*,S^*)$ is a distinguisher for $\mcP$ and $\mcQ.$
% \end{lemma} 
% %Note that from Lemma~\ref{lem:reject-sufficient-cond}, if $(i^*,S^*)$ is a distinguisher then {\EquivTester} returns {\Reject} with probability $1-o(1).$
% We will provide the proof of the above lemma in the next section. %We now complete the proof of Lemma~\ref{lem:reject-different} assuming the above lemma. 

% \subsubsection*{Proof of Lemma~\ref{lem:S(t^*)-distinguisher}}
Let for any $j \in [n]$, we define $\R(\mcD,i^*,j)$ as
\begin{align*}
&  \mcD(j) \Pr\left[\bin\left(b, \frac{\mcD(j)}{\mcD(j)+\mcD(i^*)}\right)  \le \left(\frac{1}{2}+\beta\right)b\right]   
\end{align*}
and for any subset $S \subseteq [n]$
\begin{align*}
   \R(\mcD,i^*,S) := \sum_{j \in S}\R(\mcD,i^*,j).
\end{align*}
Therefore, by the definition of tail probability,
\begin{align*}
    \TP(\mcD,i^*,S)  = \frac{\R(\mcD,i^*,S)}{\mcD(S)}.
\end{align*}
 We want to show concentration on both $\R(\mcD,i^*,S^*)$ and $\mcD(S^*),$ for all $\mcD \in \{\mcP,\mcQ\}$ which then will give us the concentration inequalities for $\TP(\mcD,i^*,S^*)$. %To this end, 
%Now we show concentration inequalities: Claim~\ref{clm:conc-Tail} bounds the values of $\R(\mcP,i^*,S^*)$ and $\R(\mcQ,i^*,S^*)$ whereas Claim~\ref{clm:conc-Mass} bounds the values of $\mcP(S^*)$ and $\mcQ(S^*)$.
\begin{claim}\label{clm:conc-Tail}
 Let $\Good_1$ be the event that for all $\mcD \in \{\mcP,\mcQ\},$ we have
       $$  |\R(\mcD,i^*,S^*) -  t^* \R(\mcD,i^*,[n])]|       \le \frac{8t^* \epsilon \sqrt{\epsilon \R(\mcD,i^*,[n])}}{c}. $$
  Then $\Pr[\Good_1] \ge 9/10.$  
\end{claim}

\begin{claim}\label{clm:conc-Mass}
   % If a set $S$ is constructed randomly  by picking each element in $[n]\setminus \{i^*\}$ with probability $t^*$ then we have
   Let $\Good_2$ be the event that all the following three conditions are satisfied 
   \begin{enumerate}
       \item $\mcP(S^*) \ge \frac{t^* \epsilon}{9}$,
       \item $\mcP(S^*) \le 200 t^*$,
       \item $\mcQ(S^*) \le 200 t^*$.
   \end{enumerate}
   
   % (i)  (ii) $\mcP(S^*) \le 200 t^*$, and (iii) $\mcQ(S^*) \le 200 t^*$, all  are satisfied. Then $\Pr[\Good_2] \ge 9/10.$  
%    all of the following inequalities hold:
% \begin{enumerate}
%  %   \item With probability at least $1-1/c$, 
% %     \begin{align}

% \item     $\mcP(S^*) \ge \frac{t^* \epsilon}{9}$, \label{item:plowerbound}
% %\end{align}
% %\item With a probability at least $1/100$
% %\begin{align}
%    \item   $\mcP(S^*) \le 200 t^*$,   \label{item:pmarkov}
% %\end{align}
% %\item with probability at least $1- \frac{1}{\log \log n},$
% and
% %\begin{align}
% \item   $\mcQ(S^*) \le 200 t^*$. \label{item:qmarkov}
% %\end{align}
% \end{enumerate} 
\end{claim}
We defer the proofs of the above two claims to the supplementary materials. Let $\Good = \Good_1 \wedge \Good_2.$ Note that by union bound, $\Pr[\Good] \geq 4/5$.

%To complete the proof, we will need the following bounds on $\mcD(S)$ for $\mcD \in \{\mcP,\mcQ\}$.

We are now ready to complete the proof of Lemma~\ref{lem:eitherProborTail}.  Assuming the event $\Good$ occurs, we now prove that $(i^*,S^*)$ is a distinguisher. 
%Since the event $\Good$ happens with at least $4/5$ probability, we are done.
For the sake of contradiction, suppose $(i^*,S^*)$ is not a distinguisher for $\mcP$ and $\mcQ.$ Then the following claim holds, the proof of which is deferred to the supplementary materials.
 \begin{claim}\label{clm:Q(S^*)>}
Assuming the event $\Good,$ and that $(i^*,S^*)$ is not a distinguisher for $\mcP$ and $\mcQ,$ we have
    $\mcQ(S^*) > \mcP(S^*)\left(1+ \frac{150 \epsilon}{c}\right)^{-1}$.
\end{claim}
  We now argue that  $\left| \frac{\mcP(i)}{\mcP(i \cup S)} - \frac{\mcQ(i)}{\mcQ(i \cup S)} \right| > 4 \gamma$, contradicting that $(i^*,S^*)$ is not a distinguisher. 
 
 % \begin{claim}
 % \label{clm:Q(S^*)> P(S^*)/}
 %     $\mcQ(S^*) > \frac{\mcP(S^*)}{1+ \frac{150 \epsilon}{c}}$.
 % \end{claim}

  %  We will first show  this implies $\mcQ(S^*) > \frac{\mcP(S^*)}{1+ \frac{150 \epsilon}{c}}$. To prove this, we need to consider two cases.

% Now we will show that $CompareMass(i^*,S^*)$ rejects. 
  % It is easy to see that    $\frac{\mcP(S^*)}{\mcP(i^*)} \le \frac{\mcQ(S^*)}{\mcQ(i^*)}$. Otherwise 
  
  If $\frac{\mcP(S^*)}{\mcP(i^*)} > \frac{\mcQ(S^*)}{\mcQ(i^*)}$ then \begin{equation}
      \frac{\mcP(S^*)}{\mcP(i^*)} > \frac{\mcQ(S^*)}{\mcQ(i^*)} > \frac{\mcP(S^*) (1 + \epsilon/4)}{\mcP(i^*) (1+ 150\epsilon/c)},
  \end{equation} (by Lemma~\ref{lem:qilesspi} and Claim~\ref{clm:Q(S^*)>}) which is a contradiction.  Hence,
\begin{equation}\label{eq:p>q}
    \frac{\mcP(S^*)}{\mcP(i^*)} \le \frac{\mcQ(S^*)}{\mcQ(i^*)}.
\end{equation}
  
  Recall that we would like to argue that $\left|\frac{\mcP(i^*)}{\mcP(i^*) + \mcP(S^*)} - \frac{\mcQ(i^*)}{\mcQ(i^*)+\mcQ(S^*)}\right| > 4 \gamma$. 
 Note that 
 \begin{align*} 
& \left|\frac{\mcP(i^*)}{\mcP(i^*) + \mcP(S^*)} -  \frac{\mcQ(i^*)}{\mcQ(i^*)+\mcQ(S^*)}\right|\\ 
& > \frac{ \frac{\mcP(S^*)}{(1 + \frac{150\epsilon}{c})\mcQ(i^*)}-\frac{ \mcP(S^*)}{\mcP(i^*)} }{\left(1+\frac{\mcP(S^*)}{\mcP(i^*)}\right)
\left(1+\frac{\mcQ(S^*)}{\mcQ(i^*)}\right)} \hspace{1em} (\text{Eq.}~\ref{eq:p>q}, \text{ Claim}~\ref{clm:Q(S^*)>})\\
& = \frac{ \frac{\mcP(S^*)}{(1 + \frac{150\epsilon}{c})}-\frac{\mcQ(i^*) \mcP(S^*)}{\mcP(i^*)} }{\left(1+\frac{\mcP(S^*)}{\mcP(i^*)}\right)\left(\mcQ(i^*)+\mcQ(S^*)\right)}\\
& > \frac{ \frac{\mcP(S^*)}{(1 + \frac{150\epsilon}{c})}-\frac{ \mcP(S^*)}{1+\epsilon/4} }{\left(1+\frac{\mcP(S^*)}{\mcP(i^*)}\right)\left(\frac{\mcP(i^*)}{1+\epsilon/4}+\mcQ(S^*)\right)} \hspace{1em}(\text{Lemma}~\ref{lem:qilesspi})\\
& > \frac{ \mcP(S^*) \epsilon }{20} \left(1+\frac{\mcP(S^*)}{\mcP(i^*)}\right)^{-1}\left(\frac{\mcP(i^*)}{1+\epsilon/4}+\mcQ(S^*)\right)^{-1}\\
%& = \frac{ \mcP(S^*) \epsilon }{100 (1+\frac{\mcP(S^*)}{\mcP(i^*)})(\frac{1}{1+\epsilon/4}+\frac{\mcQ(S^*)}{\mcP(i^*)})\mcP(i^*)}\\
& = \frac{ \epsilon }{20} \left(1+\frac{\mcP(i^*)}{\mcP(S^*)}\right)^{-1}\left(\frac{1}{1+\epsilon/4}+\frac{\mcQ(S^*)}{\mcP(i^*)}\right)^{-1}\\
& > \frac{ \epsilon }{20} \left(1+\frac{9\mcP(i^*)}{t^* \epsilon}\right)^{-1}\left(1+\frac{\mcQ(S^*)}{\mcP(i^*)}\right)^{-1} (\text{Claim}~\ref{clm:conc-Mass})\\
&  > \frac{ \epsilon }{20} \left(1+\frac{9 t^* \epsilon^3}{c^2 t^* \epsilon}\right)^{-1}\left(1+\frac{\mcQ(S^*)}{\mcP(i^*)}\right)^{-1} \hspace{1em}(\text{Eq.}~\ref{eq:t^*-P(i^*)})\\
& > \frac{ \epsilon }{40} \left(1+\frac{\mcQ(S^*)}{\mcP(i^*)}\right)^{-1}\\
& > \frac{ \epsilon }{40} \left(1+\frac{200 t^* }{ \mcP(i^*)}\right)^{-1} \hspace{1em}(\text{Claim}~\ref{clm:conc-Mass})\\
& > \frac{ \epsilon }{40} \left(1+\frac{400 t^* c^2 }{ t^* \epsilon^3}\right)^{-1} \hspace{1em} (\text{Eq.}~\ref{eq:t^*-P(i^*)})\\
%&  > \frac{ \epsilon }{200 (1+\frac{200 t^* }{ \mcP(i^*)})}(\text{from} \thinspace (\ref{eq:qmarkov}))\\
& > \frac{ \epsilon }{40 \left(\frac{401  c^2 }{  \epsilon^3}\right)} > \frac{\epsilon^{4}}{ 16040 c^2} >4 \frac{\epsilon^{4}}{ L} = 4 \gamma. & \qedhere
\end{align*}
% which is a contradiction.
\end{proof}

\section{Conclusion}
\label{sec:conclusion}

We considered the problem of equivalence testing of two distributions (over $[n]$) in the conditional sampling model. We presented a simple algorithm with sample complexity $\tilde{O}(\log n)$. While our algorithm is not fully non-adaptive, it is only one-round adaptive. This shows that even a limited amount of adaptiveness can help to significantly reduce the sample/query complexity. Our algorithm can also be modified slightly to obtain a fully adaptive algorithm with sample complexity $\tilde{O}(\log\log n)$, matching the best bound in this setting.

One limitation of our algorithm is the presence of large constants and a worsened dependency on the parameter \(\epsilon\) compared to the previous algorithm by~\cite{kamath2019anaconda}. Investigating methods to reduce this dependency on \(\epsilon\) while maintaining the \(\tilde{O}(\log n)\) dependency with respect to the parameter \(n\) poses an intriguing open direction of research.

\section{Acknowledgements}
 %All acknowledgments go at the end of the paper, including thanks to reviewers who gave useful comments, to colleagues who contributed to the ideas, and to funding agencies and corporate sponsors that provided financial support. 
%To preserve the anonymity, please include acknowledgments \emph{only} in the camera-ready papers.
%We thank the anonymous reviewers for their useful comments.
We thank the anonymous reviewers for their useful comments. D. Chakraborty is supported in part by an MoE AcRF Tier 2 grant (MOE-T2EP20221-0009), an MoE AcRF Tier 1 grant (T1 251RES2303), and a Google South \& South-East Asia Research Award. K. S. Meel is supported in part by National Research Foundation Singapore under its NRF Fellowship Programme [NRF-NRFFAI1-2019-0004 ], Ministry of Education Singapore Tier 2 grant MOE-T2EP20121-0011, and Ministry of Education Singapore Tier 1 Grant [R-252-000-B59-114].

\bibliography{references}
\newpage 
\appendix
\section{MISSING PROOFS}

\begin{proof}[Proof of Lemma~\ref{lem:estimatorD(i)/D(i)+D(S)}]
Note that for any $k \in [m]$, $\mathbbm{1}_{j_k =i}$ is a Bernoulli random variable such that $\Pr[\mathbbm{1}_{j_k =i} =1] = \Pr[j_k = i] = \frac{\mcD(i)}{\mcD(i \cup S)}$. Also, note that  $\mathbbm{1}_{j_k =i}$  are iid for all $k\in [m]$. By additive Chernoff bound (Lemma~\ref{lem:additiveChernoffBound}), the value $\mathsf{ep} = \EstProb(\mcD,i,S,m)$ returned by  the estimator  satisfies:
\begin{align*}
%\label{eq:Chernoff-CompareMass}
\Pr\left[\left|\mathsf{ep} - \frac{\mcD(i)}{\mcD(i \cup S)} \right| > \gamma\right] \le e^{-\gamma^2 m}.
\end{align*}
    
\end{proof}

\begin{proof}[Proof of Lemma~\ref{lem:estimatorTailProb}]
For any $k \in [m]$, we have
\begin{align*}
 &   \Pr[Z_k = 1] \\
  &  = \sum_{j \in S} \Pr[j_k = j] \Pr\left[\sum_{\ell \in [b]}\mathbbm{1}_{y_\ell = j} \le \left(1/2+\beta \right)b\right]\\
  &   =  \sum_{j \in S} \frac{\mcD(j)}{\mcD(S)} \Pr\left[\bin\left(b,\frac{\mcD(j)}{\mcD(i)+\mcD(j)}\right) \le (1/2+\beta)b\right]\\
  &    = \TP(\mcD,i,S,\beta,b).
\end{align*}
Therefore, by additive Chernoff bound (Lemma~\ref{lem:additiveChernoffBound}), the value $\mathsf{et} = \EstTail \left(\mcD,i,S,\beta,b,m \right)$ returned by the algorithm satisfies
\begin{align*}
%\label{eq:Chernoff-CompareProb}
\Pr[\left|\mathsf{et} - \TP \left(\mcD,i,S,\beta,b \right) \right| > \gamma] \le e^{-\gamma^2 m}.
\end{align*}

 %   Fix any $X_k$ for $k \in [N]$. We have $\Pr[X_k = j] = \frac{D(j)}{D(S)}$ for any $j \in S$. Also for any $\ell \in [B]$, $\Pr[1_{Y_\ell = X_k}] =\Pr[Y_\ell = X_k] = \frac{D(X_k)}{D(i)+D(X_k)}.$  Thus  $\sum_{\ell \in [B]}1_{Y_\ell = X_k}$ is distributed as binomial random variable $Bin(B,\frac{D(X_k)}{D(i)+D(X_k)})$ and the random variable $Z_k$ is a Bernoulli random variable with success probability $\Pr[Bin(B,\frac{D(X_k)}{D(i)+D(X_k)}) \le (1/2+\beta)B]$.
\end{proof}

To prove Claims~\ref{clm:conc-Tail}, \ref{clm:conc-Mass} and \ref{clm:Q(S^*)>}, we will use the following concentration inequality that directly follows from Bernstein's concentration inequality.
\begin{lemma}~\cite{falahatgar2015faster}
\label{lem:bernstein}
Consider a set $G$ and a function $r:G \rightarrow \mathbb{R}_{\ge 0}$ such that $\max_{j\in G} r(j) \le  r_{max}$. Consider set $S$ formed by selecting each element from $G$ independently and uniformly randomly with probability $r_0$, then $\mathbb{E}[r(S)]= r_0 r(G)$ and with probability at least $1-2 \lambda$,
\[
|r(S) - \mathbb{E}[r(S)]| \le \sqrt{2r_0 r_{max}r(G) \log \frac{1}{\lambda}} + r_{max} \log \frac{1}{\lambda}.
\]
\end{lemma}

We now first state lower bounds for $ \R(\mcP,i^*,[n])$ and $\R(\mcQ,i^*,[n])$ in the following claim.

 \begin{claim}
\label{clm:taillarge}
\begin{align}
\label{eq:tail_1large}
    \R(\mcP,i^*,[n]) \ge \frac{\epsilon}{9}
\end{align}
\begin{align}\label{eq:tail_2large}
      \R(\mcQ,i^*,[n]) \ge \frac{\epsilon}{10}
\end{align}
\end{claim}

\begin{proof}
 Note that if $\mcP(j) \le \mcP(i^*)$ then by additive Chernoff bound, $\Pr\left[\bin\left(B,\frac{\mcP(j)}{\mcP(j)+\mcP(i^*)}\right) \le (\frac{1}{2}+\beta) B\right] \ge 1 - \frac{1}{e^{\beta^2 B}} \ge 1 - \frac{1}{e^{100}}$. Hence, we have
\begin{align*}
     \R(\mcP,i^*,[n])  & =    \sum_{j \in [n]} \mcP(j) \Pr\left[\bin\left(b,\frac{\mcP(j)}{\mcP(j)+\mcP(i^*)}\right)    \le (\frac{1}{2}+\beta) b\right] \\
 &     \ge (1 - \frac{1}{e^{100}})\sum_{j: \mcP(j) \le \mcP(i^*)}\mcP(j) \ge \frac{\epsilon}{9} %\ge \frac{\epsilon}{8}
\end{align*}
where the last inequality is because of item \ref{item:taili^*bound} of Lemma~\ref{lem:qilesspi}.
Now we have,
\begin{align*}
   \R(\mcQ,i^*,[n])  & =     \TP(\mcQ,i^*,[n]) \\
&   \ge  (\TP(\mcP,i^*,[n]) - 4 \gamma )    ( \text{ from inequality (\ref{eq:zpclosezq}) })\\
& =   (\R(\mcP,i^*,[n]) - 4 \gamma ) \\
&  \ge \epsilon/10  (\text{as $\gamma = \epsilon^4/L$ })
\end{align*}

\end{proof}

Now we prove the Claims~\ref{clm:conc-Tail}, \ref{clm:conc-Mass} and \ref{clm:Q(S^*)>}.
\subsection*{Proof of Claim~\ref{clm:conc-Tail}}

%We are interested in the high-probability bounds on the value of $TailProb(D,i^*,S)$ when $S$ is constructed by picking each element in $[n]\setminus \{i^*\}$ with probability $t^*$. Since, $TailProb(D,i^*,S)$ is the ratio of $R(D,i^*,S)$ and $D(S)$, we first determine the expected values of $R(D,i^*,S)$ and $D(S)$ and then give concentration inequalities on them separately. For the same, we will use the following concentration result which directly follows from the well known Bernstein inequality.

%\paragraph*{Proof of Claim~\ref{clm:conc-Tail}:}
Now we provide the proof of our Claim~\ref{clm:conc-Tail}. 
Consider an indicator random variable $\mathbbm{1}_{j \in S^*}$ for each $j \in [n]$ that takes the value $1$ if $j$ is included in $S^*$ and $0$ otherwise. We have 
 \begin{align*}%\label{eq:expected-TailProb}
         \E[\R(\mcD,i^*,S)] =        \sum_{j \in [n]} \E[\mathbbm{1}_{j \in S^*}]  \R(\mcD,i^*,j) = t^*  \R(\mcD,i^*,[n])%  (1-D(i^*)) TailProb(D,i^*,[n]\setminus \{i^*\})
    \end{align*}
    Note that if $\mcD(j) \ge \frac{3}{2} \mcD(i^*)$ then by additive Chernoff bound, we have  $ \bin(b,\frac{\mcD(j)}{\mcD(j)+\mcD(i^*)} ) \le \frac{1}{2} + \beta$ with probability at most $\frac{1}{e^{(3/5 - 1/2 - \beta)^2 b}} = 1/e^{100}$. Therefore, $\max_{j}\R(\mcD,i^*,j) \le 2 \mcD(i^*)$. %(using inequality (\ref{eq:pismall})). 

To apply Lemma~\ref{lem:bernstein}, we have set $G = [n]$,  $r_j = \R(\mcD,i^*,j)$ for all $j \in [n]$, $r_0 = t^*$ and $r_{max} = 2 \mcD(i^*)$ and $\lambda = \frac{1}{c}$. Therefore, with probability at least $1-1/\lambda = 1- 1/c,$

%(with $r_0 = t, r_{max} = 2D(i^*), r(G) = R_D(i^*,[n]\setminus \{i^*\}), \delta = \frac{1}{2 (\log  \log n)^2}]$, we have with probability at least
\begin{align*}
    |\R(\mcD,i^*,S^*) - t^* \R(\mcD,i^*,[n])|  &  \le   \sqrt{2t^* (2 \mcD(i^*)) \R(\mcD,i^*,[n]) ( \log c )} + 2 \mcD(i^*) ( \log c)  \\
 &  \le   \frac{8 t^* \epsilon \sqrt{\epsilon   \R(\mcD,i^*,[n])}}{c}
\end{align*}
where the last inequality is clear from the fact that $ \mcQ(i^*) < \mcP(i^*)$ (Lemma~\ref{lem:qilesspi}) , $\mcP(i^*) \le t^* \epsilon^3/c^2$ (inequality (\ref{eq:t^*-P(i^*)})), $\mcP(i^*) \le 32\epsilon^3/L$ (inequality (\ref{eq:pismall})) and $c=1000, L = 10^{15}.$
%\end{proof}

\subsection*{Proof of Claim~\ref{clm:conc-Mass}}
 Consider the set $G = \{j \in [n]:\mcP(j) \le \mcP(i^*)\}.$
%    (recall that $Tail_D(i,[n],\infty) \ge \frac{\epsilon}{8}$ and so $Tail_D(i,[n],\infty) - \mcP(i^*) \ge \frac{\epsilon}{8.1}$).
Further, let $S^*_G = S^* \cap G$. Obviously, $\mcP(S^*) \ge \mcP(S^*_G)$. Further, $\E[\mcP(S^*_G)] = t^* \cdot (\sum_{j: \mcP(j) \le \mcP(i^*)} \mcP(j)) \ge 3 t^* \epsilon/10$ (from Lemma~\ref{lem:qilesspi}).

Applying Lemma~\ref{lem:bernstein} to the set $G$ with $r_0 = t^*, r_{max} = \mcP(i^*), r(G) = \mcP(G) \le 1$ and $\lambda = \frac{1}{c}$, we have $\Pr[\mcP(S_G) < 3 t^* \epsilon/10 - 4 t^* \epsilon \sqrt{\epsilon }/ c] <$ $\frac{1}{c}$. Note that $3 t^* \epsilon/10 - 4 t^* \epsilon \sqrt{\epsilon }/ c > \epsilon/9.$
Therefore,% with probability at most $\frac{2}{\log \log n},$ we have $p(S) \not \in [ \frac{t \epsilon}{8}, t \log \log n].$ 
\begin{align*}
  \Pr[\mcP(S^*) < \frac{t \epsilon}{9}] <   \Pr[\mcP(S^*_G) < \frac{t \epsilon}{9}] <\frac{1}{c}
\end{align*}

Note that $\E[\mcP(S^*)] = \E[\mcQ(S^*)] = t^*$. Therefore, by Marhov's inequality, we have $\Pr[\mcP(S^*) > 200 t^*] < 1/200$ and  $\Pr[\mcQ(S^*) > 200 t^*] < 1/200.$ By union bound, our claim holds.

\subsection*{Proof of Claim~\ref{clm:Q(S^*)>}}
First, we show the following claim, which is a corollary of Claim~\ref{clm:conc-Tail}.
\begin{claim}\label{ref:clm:R1closerR2}
Assuming the event $\Good,$ we have
    \begin{align}\label{eq:R1closerR2}
  |\R(\mcP,i^*,S^*) - \R(\mcQ,i^*,S^*)| \le  \frac{20 t^* \epsilon \sqrt{\epsilon \min\{\R(\mcP,i^*,[n]),\R(\mcQ,i^*,[n])\}}}{c}.
\end{align} 
\end{claim}

\begin{proof}
By triangle inequality, we have
     \begin{align*}
     |\R(\mcP,i^*,S^*) - \R(\mcQ,i^*,S^*)| &  \le |\R(\mcP,i^*,S^*) - t^* \R(\mcP,i^*,[n])| +    |\R(\mcQ,i^*,S^*) - t^* \R(\mcQ,i^*,[n])|  + \\
     & t^* |\R(\mcP,i^*,[n])    - \R(\mcQ,i^*,[n])|\\
&      \le \frac{8 t^* \epsilon \sqrt{\epsilon \R(\mcP,i^*,[n])}}{c} +     \frac{8 t^* \epsilon \sqrt{\epsilon \R(\mcQ,i^*,[n])}}{c}+ 4t^* \gamma.\\
%  &   \le \frac{20 t^* \epsilon \sqrt{\epsilon \min\{\R(\mcP,i^*,[n]\setminus \{i^*\}),\R(\mcQ,i^*,[n]\setminus \{i^*\})\}}}{c}
   \end{align*} 
The  last inequality implies from Claim~\ref{clm:conc-Tail}, the inequality (\ref{eq:zpclosezq})  and that $\TP(\mcP,i^*,[n]) = \R(\mcP,i^*,[n]).$   Further, we have 
\begin{align*}
 |\frac{8 t^* \epsilon \sqrt{\epsilon \R(\mcP,i^*,[n])}}{c} - \frac{8 t^* \epsilon \sqrt{\epsilon \R(\mcQ,i^*,[n])}}{c}| &=  \frac{8t^*\epsilon \sqrt{\epsilon}}{c} |\sqrt{\R(\mcP,i^*,[n])}-\sqrt{\R(\mcQ,i^*,[n])}| \\
& \le  (8t^*\epsilon^{3/2}/c) \frac{|\R(\mcP,i^*,[n]) -\R(\mcQ,i^*,[n])|}{\sqrt{\R(\mcP,i^*,[n])}+\sqrt{\R(\mcQ,i^*,[n])}} \\
& \le (8t^*\epsilon^{3/2}/c) \frac{12 \gamma}{2 \sqrt{\epsilon}} \\
&\le  50t^* \epsilon \gamma/c.
\end{align*}
The second last inequality is because of inequalities (\ref{eq:tail_1large}) and (\ref{eq:tail_2large}).

Therefore, we have 
\begin{align*}
& |\R(\mcP,i^*,S^*) - \R(\mcQ,i^*,S^*)|     \\
&  \le \frac{8 t^* \epsilon \sqrt{\epsilon \R(\mcP,i^*,[n])}}{c} +     \frac{8 t^* \epsilon \sqrt{\epsilon \R(\mcQ,i^*,[n])}}{c}+ 4t^* \gamma \\
& \le  \frac{16 t^* \epsilon \sqrt{\epsilon \min\{\R(\mcP,i^*,[n]),\R(\mcQ,i^*,[n])\}}}{c}   + |\frac{8 t^* \epsilon \sqrt{\epsilon \R(\mcP,i^*,[n])}}{c}  - \frac{8 t^* \epsilon \sqrt{\epsilon \R(\mcQ,i^*,[n])}}{c}| + 4t^* \gamma \\
& \le  \frac{16 t^* \epsilon \sqrt{\epsilon \min\{\R(\mcP,i^*,[n]),\R(\mcQ,i^*,[n])\}}}{c} + (8t^*\epsilon^{3/2}/c) \frac{12 \gamma}{2 \sqrt{\epsilon}} + 4t^* \gamma \\
&\le  \frac{20 t^* \epsilon \sqrt{\epsilon \min\{\R(\mcP,i^*,[n]),\R(\mcQ,i^*,[n])\}}}{c}.
\end{align*}

The last inequality holds because  $\frac{16 t^* \epsilon \sqrt{\epsilon \R(\mcP,i^*,[n])}}{c}$ is  at least $16 t^* \epsilon \sqrt{\epsilon \epsilon/10}/c \ge t^* \epsilon^2/c$ 
whereas \\ $50t^* \epsilon \gamma/c + 4t^* \gamma \le 100 t^* \gamma = 100 t^* \epsilon^4/L.$

%\begin{align*}
%&     |\R(\mcP,i^*,S^*) - \R(\mcQ,i^*,S^*)| \le \\
%&      \frac{8 t^* \epsilon \sqrt{\epsilon R(\mcP,i^*,[n])}}{c} +     \frac{8 t^* \epsilon \sqrt{\epsilon R(\mcQ,i^*,[n])}}{c}+ 4t^* \gamma\\
%& \le \frac{8 t^* \epsilon \sqrt{\epsilon \min\{\R(\mcP,i^*,[n]),\R(\mcQ,i^*,[n])\}}}{c} + |\frac{8 t^* \epsilon \sqrt{\epsilon R(\mcP,i^*,[n])}}{c} - \frac{8 t^* \epsilon \sqrt{\epsilon R(\mcQ,i^*,[n])}}{c}|+ 4t^* \gamma
%\end{align*}

%The last inequality comes from the fact that $|R(\mcP,i^*,[n]\setminus \{i^*\}) - R(\mcQ,i^*,[n]\setminus \{i^*\})| \le O(\gamma)$ ( which comes from the fact that $|TailProb(\mcP,i^*,[n]\setminus \{i^*\}) - TailProb(\mcQ,i^*,[n]\setminus \{i^*\})| \le 4\gamma$ and that $\gamma = O(\epsilon^{2})$ while  $TailProb(\mcQ,i^*,[n]\setminus \{i^*\}),TailProb(\mcP,i^*,[n]\setminus \{i^*\}) \ge \epsilon/10.$)
\end{proof}

%    Assuming $CompareTail(i^*,S^*)$ does not reject we will show that $CompareProb(i^*,S^*)$ rejects.  Since $CompareTail(i^*,S^*)$ does not reject so we have  $4 \gamma \ge |\frac{R(\mcQ,i^*,S^*)}{\mcQ(S^*)} - \frac{R(\mcP,i^*,S^*)}{\mcP(S^*)}|$.
Now we proceed with showing the proof of Claim~\ref{clm:Q(S^*)>}. Recall that $\TP(\mcD,i^*,S^*) = \R(\mcD,i^*,S^*)/\mcD(S^*).$
We consider two cases.
 \begin{enumerate}
        \item   $ \frac{\R(\mcP,i^*,S^*)}{\mcP(S^*)} < \frac{\R(\mcQ,i^*,S^*)}{\mcQ(S^*)}$.% \Recall that $\delta_4 = \frac{t \epsilon^2}{16 \log \log n}$.
 
 By assumption that $(i^*,S^*)$ is not a distinguisher, we have
\begin{align*}
     4 \gamma & \ge \frac{\R(\mcQ,i^*,S^*)}{\mcQ(S^*)} - \frac{\R(\mcP,i^*,S^*)}{\mcP(S^*)}\\
    & \ge \R(\mcP,i^*,S^*) (\frac{1}{\mcQ(S^*)} - \frac{1}{\mcP(S^*)}) -    \frac{20 t^* \epsilon \sqrt{\epsilon \R(\mcP,i^*,[n])}}{c \cdot \mcQ(S^*)}(\text{from} (\ref{eq:R1closerR2}))
 \end{align*}
 This implies that 
 \begin{align*}
     \R(\mcP,i^*,S^*) (\frac{1}{\mcQ(S^*)} - \frac{1}{\mcP(S^*)}) &  \le 4\gamma + \frac{20 t^* \epsilon \sqrt{\epsilon \R(\mcP,i^*,[n])}}{c \cdot \mcQ(S^*)}  \\
    & \le \frac{25 t^* \epsilon \sqrt{\epsilon \R(\mcP,i^*,[n]\setminus \{i^*\})}}{c \cdot \mcQ(S^*)}
   \end{align*}
   The last inequality follows because $4 \gamma = 4 \epsilon^4/L$ is small compared to $ \frac{20 t^* \epsilon \sqrt{\epsilon \R(\mcP,i^*,[n])}}{c \cdot \mcQ(S^*)} \ge \frac{20 t^* \epsilon \sqrt{\epsilon \cdot  \epsilon/9}}{c 200 t^*} \ge \epsilon^2/200c.$ Here, we used  Claim~\ref{clm:conc-Mass} and 
 inequality (\ref{eq:tail_1large}). Now,
   \begin{align*}
  &  \R(\mcP,i^*,S^*) (\frac{1}{\mcQ(S^*)} - \frac{1}{\mcP(S^*)})  \le \frac{25 t^* \epsilon \sqrt{\epsilon \R(\mcP,i^*,[n])}}{c \cdot \mcQ(S^*)}\\
     \implies & \mcP(S^*)-\mcQ(S^*) < \frac{25 t \epsilon \sqrt{\epsilon \R(\mcP,i^*,[n])} \mcP(S^*)}{c \cdot \R(\mcP,i^*,S^*) }\\
     \implies & \mcP(S^*)-\mcQ(S^*) < \frac{30 \epsilon \sqrt{\frac{\epsilon}{ \R(\mcP,i^*,[n])}} \mcP(S^*)}{c   }
   \end{align*} 
   The last inequality comes from fact that $\R(\mcP,i^*,S^*) \ge t^* \R(\mcP,i^*,[n]) (1 - 8/c)$(\text{from Claim~} \ref{clm:conc-Tail}). 
 Finally,  since $\R(\mcP,i^*,[n]) \ge \epsilon/9$(\text{from} (\ref{eq:tail_1large})), we have $\mcP(S^*)-\mcQ(S^*) < 90 \epsilon \mcP(S^*)/c$ and hence the claim follows.
   
\item  $\frac{\R(\mcP,i^*,S^*)}{\mcP(S^*)} > \frac{\R(\mcQ,i^*,S^*)}{\mcQ(S^*)}.$ 

From (\ref{eq:R1closerR2}), we have
 \begin{align*}
        & 0 < \frac{\R(\mcP,i^*,S^*)}{\mcP(S^*)} - \frac{\R(\mcQ,i^*,S^*)}{\mcQ(S^*)}  < \frac{\R(\mcQ,i^*,S^*)}{\mcP(S^*)} -  \frac{\R(\mcQ,i^*,S^*)}{\mcQ(S^*)}  + \frac{20 t^* \epsilon \sqrt{\epsilon  \R(\mcQ,i^*,[n])}}{c} 
  \end{align*}
  This implies 
  \begin{align*}
        &  \R(\mcQ,i^*,S^*)(\frac{\mcP(S^*) - \mcQ(S^*)}{\mcP(S^*)\mcQ(S^*)})  <  \frac{20 t^* \epsilon \sqrt{\epsilon \R(\mcQ,i^*,[n])}}{c \cdot \mcP(S^*)}.
      \end{align*}
      Hence, $(\frac{\mcP(S^*) - \mcQ(S^*)}{\mcQ(S^*)}) < \frac{20 t^* \epsilon \sqrt{\epsilon \R(\mcQ,i^*,[n])}}{c \cdot \R(\mcQ,i^*,S^*)} $  which is at most 
      $\frac{25 t^* \epsilon \sqrt{\epsilon \R(\mcQ,i^*,[n])}}{c  t^* \R(\mcQ,i^*,[n]) }$ (from Claim~ \ref{clm:conc-Tail})).  Finally,
      $\frac{25 t^* \epsilon \sqrt{\epsilon \R(\mcQ,i^*,[n])}}{c  t^* \R(\mcQ,i^*,[n]) } = \frac{25 \epsilon\sqrt{\frac{\epsilon}{\R(\mcQ,i^*,[n])}}}{ c} \le \frac{90 \epsilon}{ c} $ (from inequality  (\ref{eq:tail_2large})). Thus, we have  $(\frac{\mcP(S^*) - \mcQ(S^*)}{\mcQ(S^*)}) \le \frac{90 \epsilon}{ c}$ and this directly  implies the claim.
    \end{enumerate}

\section{An $\Tilde{O}(\log \log n)$-query fully adaptive algorithm}
Our algorithm can also be modified slightly to obtain a fully adaptive algorithm with sample complexity $\tilde{O}(\log\log n)$. This matches the best-known bound in this setting  by~\cite{falahatgar2015faster}. In the original formulation, our algorithm sequentially examines all $\log n$ possible values of $t$ to find a particular value, $t^*$. At $t^*$, one of our subroutines—either \EstProb{} or \EstTail{}—will \Reject{} if the input distributions significantly differ. Employing a binary search for $t^*$  reduces the number of queries to $\tilde{O}(\log \log n)$. However, this process requires adaptivity at each iteration.

\end{document}